\documentclass[%
 reprint,
 amsmath,amssymb,
 aps,
]{revtex4-2}

\usepackage{graphicx}% Include figure files
\usepackage{dcolumn}% Align table columns on decimal point
\usepackage{bm}% bold math
\usepackage{comment}
\usepackage{subfig,epsfig,tikz,float}
\usepackage{booktabs,multirow,tabularx,array} 
\usepackage{natbib}
\usepackage[T1]{fontenc}
\newcommand{\ditto}[1][.4pt]{\textquotedbl}

\begin{document}

\title{Fourier-space generalized magneto-optical ellipsometry}% Force line breaks with \\

\author{Miguel A. Cascales Sandoval$^{1}$, A. Hierro-Rodríguez$^{*2,3}$, D. Sanz-Hernández$^{4}$, L. Skoric$^{5}$,\\ C. N. Christensen$^{5}$, C. Donnelly$^{6}$ and  A. Fernández-Pacheco$^{*1,7}$}
\bigskip
\affiliation{*Corresponding author e-mails: hierroaurelio@uniovi.es, amaliofp@unizar.es.}
\affiliation{$^{1}$University of Glasgow, Glasgow G12 8QQ, UK}
\affiliation{$^{2}$Departamento de Física, Universidad de Oviedo, 33007, Oviedo, Spain}
\affiliation{$^{3}$CINN (CSIC – Universidad de Oviedo), 33940, El Entrego, Spain}
\affiliation{$^{4}$Unité Mixte de Physique, CNRS, Thales, Université Paris-Saclay, 91767, Palaiseau, France}
\affiliation{$^{5}$University of Cambridge, Cambridge CB3 0HE, UK}
\affiliation{$^{6}$Max Planck Institute for Chemical Physics of Solids, 01187 Dresden, Germany}
\affiliation{$^{7}$Instituto de Nanociencia y Materiales de Aragón, CSIC-Universidad de Zaragoza, 50009 Zaragoza, Spain}

%\date{\today}

\begin{abstract}
The magneto-optical Kerr effect (MOKE) is a widely used lab-based technique for the study of thin films and nanostructures, providing magnetic characterization with good spatial and temporal resolutions. Due to the complex coupling of light with a magnetic sample, conventional MOKE magnetometers normally work by selecting a small range of incident wave-vector values, focusing the incident light beam to a small spot, and recording the reflected intensity at that angular range by means of photodetectors. This generally provides signals proportional to a mixture of magnetization components, requiring additional methodologies
for full vectorial magnetic characterization.

Here, we computationally investigate a Fourier-space MOKE setup, where a focused beam ellipsometer using high numerical aperture optics and a camera detector is employed to simultaneously map the intensity distribution for a wide range of incident and reflected wave-vectors. We employ circularly incident polarized light and no analyzing optics, in combination with a fitting procedure of the light intensity maps to the analytical expression of the Kerr effect under linear approximation. In this way, we are able to retrieve the three unknown components of the magnetization vector as well as the material’s optical and magneto-optical constants with high accuracy and short acquisition times, with the possibility of single shot measurements. Fourier MOKE is thus proposed as a powerful method to perform generalized magneto-optical ellipsometry for a wide range of magnetic materials and devices.
\end{abstract}

\maketitle

%\tableofcontents

\section{Introduction}

Many novel spintronic systems currently under investigation present complex magnetic states with three-dimensional spatial dependence \cite{gobel2021beyond}, making necessary the development of robust characterization techniques able to perform vector magnetometry or magnetic microscopy \cite{fernandez2017three,donnelly2020imaging} where all three components of the magnetization can be determined.

The magneto-optical Kerr effect (MOKE) is a flexible, accessible, lab-based technique widely used for the characterization of magnetic materials. MOKE has excellent magnetic sensitivity, as well as very good spatial and temporal resolutions \cite{mccord2015progress,kleemann2007perspective}, and is usually employed in either MOKE magnetometers or Kerr microscopes. In MOKE magnetometers, a focused laser is used to probe a small area \cite{teixeira2011versatile,jimenez2014vectorial,allwood2003magneto} of the sample, which also allows to perform scanning Kerr microscopy using either mobile stages or beams \cite{flajvsman2016high}. A Kerr microscope, on the contrary, uses a modified full field-of-view polarization microscope for the direct imaging of magnetic domains \cite{mccord2015progress,soldatov2017selective,choudhary2022grain,soldatov2017advanced,idigoras2010kerr}. 

Despite its popularity and the relative ease of use of MOKE systems, the Kerr effect, that describes the coupling of light and a magnetic material under reflection, is far from trivial. Even in the simplest case, where only the linear Kerr effect is considered, the complex nature of the refractive index $n$ and Voigt constant $Q$ (the latter describing the magneto-optical response of the material), added to the tensorial relationship between electric field and the magnetization, gives rise to a mixture of Kerr signals coming from different components of the magnetization. This generally prevents a quantitative solution being obtained for the evolution of the three components of the magnetization under external stimuli, typically magnetic fields.

Over the years, different approaches have been followed to perform vectorial MOKE, in order to obtain a quantitative picture of the 3D evolution of the magnetization vector field during a hysteresis loop. The quantitative determination of magnetic and optical properties of a material is highly attractive, however generally requires a combination of measurements taken under different magneto-optical conditions. For instance, in cases where only in-plane (IP) magnetization is present, the use of polarizing optics and the fact that longitudinal and transverse Kerr effects manifest differently, either as a rotation/ellipticity or as a change in relative intensity, makes it possible to perform vector magnetometry \cite{jimenez2014vectorial}. In cases where the magnetization evolves throughout the whole space, generalized vector magnetometry becomes necessary, which may be achieved by exploiting symmetry arguments of the Kerr effect, \textit{e.g.} performing differential analysis of the signal at two symmetric incident wave-vectors ($\vec{k}$), which allows the decoupling of the IP and out-of plane (OOP) components \cite{ding2000experimental}. This can be also achieved by advanced light modulation schemes under reasonable approximations of the Kerr signal \cite{vavassori2000polarization}. Finally, generalized magneto-optical ellipsometry (GME) is a vector magnetometry technique where not only the components of the magnetization, but all optical and magneto-optical constants of a system are obtained. This approach gives a complete quantitative magneto-optical characterization of a material, as initially demonstrated by \textit{Berger et al.} \cite{berger1997generalized}. So far, the implementation of GME consists of measuring systematically the reflected signal as a function of  different polarizer-analyzer angles, which are fitted to the Kerr effect formalism under linear approximation \cite{berger1997generalized}. A second approach, where a rotating quarter wave-plate is added before the analyzer is also reported on the literature \cite{oblak2020ultrasensitive}. Using either of these configurations, GME has been employed for vector magnetometry and spectroscopy in depth studies \cite{neuber2003temperature,mok2011vector,arregi2015strain,berger1999quantitative}.

In this work we computationally investigate a new Fourier-space based MOKE magnetometry technique for GME. For this, a setup based on focused beam ellipsometers \cite{ye2007development,ye2008ellipsometric,lee2020co} is proposed. Here, light is sent and collected from a metallic magnetic material at a wide range of $\vec{k}$ using high-numerical aperture optics. The reflected beam is collected by means of an area detector, providing a map of recorded intensities as a function of different $\vec{k}$ values. This contrasts with standard MOKE approaches where the $\vec{k}$ employed are not spatially resolved. We thus obtain a magneto-optical intensity map in the Fourier-space, corresponding to a 3D magnetic state in the sample. The resulting intensity pattern is fitted to the analytical expression of the linear Kerr effect \cite{berger1997generalized}, which allows us to obtain the three components of the magnetization, as well as the optical and magneto-optical constants of the material. We find that this technique works best when employing circularly incident polarized light without analyzing optics.

We showcase our newly proposed technique "Fourier MOKE", presented here by studying how our fitting protocol performs in retrieving the GME parameters for various intensity maps, computationally generated to represent different magnetic materials and single magnetic states. These maps are also modified to include different sources of noise and camera characteristics, which allows us to test the robustness of our procedure under more realistic experimental conditions. We observe that the technique provides a robust reconstruction of the three components of the magnetization, as well as optical and magneto-optical constants. Remarkably, we find that similar precision to conventional GME \cite{berger1997generalized} can be obtained with relatively low statistics. Furthermore, with single shot measurements it is possible to obtain the magnetisation vector orientation with an angular error of $5^{\circ}$, making possible the study of \textit{e.g.} domain wall dynamic and stochastic processes with high precision. Finally we apply this procedure to the case of a magnetic loop, where the magnetization vector evolves over time along the whole space. We exploit the fact that all the maps in a loop correspond to the same material, by employing a self-consistent algorithm which simultaneously fits all maps corresponding to the different magnetic states in the loop. We further demonstrate how this additional layer of complexity in the data analysis protocol improves the accuracy in the determination of optical and magneto-optical constants.  We thus propose the Fourier MOKE technique as a simple but powerful method for the advanced characterization of magnetic systems with three-dimensional magnetic states, extending the capabilities of focused beam ellipsometers which have been previously used solely for non-magnetic materials.

\section{Theory}

\begin{figure}[h]
\centering
\includegraphics[scale=0.7]{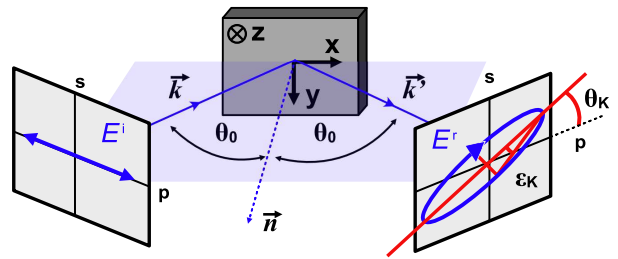}
\caption{\label{fig:moke_intro} Interaction of linearly polarized incident light $E^{i}$ at incidence angle $\theta_{0}$ with a magnetic medium acquiring Kerr rotation $\theta_{K}$ and Kerr ellipticity $\epsilon_{K}$.}
\end{figure}

Although initially observed by the Scottish physicist John Kerr in 1876 \cite{protopopov2014practical}, MOKE was not successfully explained until the development of quantum mechanical theory. Hulme, Kittel and Argyres related a change in refractive index for both circular eigenmodes of light propagation to changes in the electron’s spin dependent part of the wavefunction due to the spin-orbit interaction, which average to non-zero in a magnetic material \cite{qiu1999surface}. This difference in refractive index yields different propagation velocities (Magnetic Circular Birefringence, MCB) and different absorptions (Magnetic Circular Dichroism, MCD) for both circular eigenmodes. MCB and MCD respectively cause Kerr rotation ($\theta_{K}$) and Kerr ellipticity ($\epsilon_{K}$) on linearly polarized light, since this can be expressed as the superposition of both circular eigenmodes (figure \ref{fig:moke_intro}).

Solving Maxwell’s equations and applying the appropriate boundary conditions yields Fresnel’s reflection coefficients, $r_{ij}$, which represent the ratio of reflected ($E^{r}$) i-polarized electric field component with respect to the incident ($E^{i}$) j-polarized component. These are given in expressions \ref{eq:rss}-\ref{eq:rps} when utilizing the sign convention from \cite{wu2000optical,loughran2018enhancing,hubert2008magnetic} and expressed in the $p-s$ polarization basis, respectively denoting the directions parallel and perpendicular to the optical plane. The frame of reference (FOR) basis for the magnetization vector ($\vec{m}$) is given by the optical plane and sample relative orientations \cite{wu2000optical}, as shown in figure \ref{fig:moke_intro}. The $\vec{x}$ direction lies on the optical plane and orthogonal to the surface's normal vector, the $\vec{y}$ direction is orthogonal to both the optical plane and the surface's normal, whereas the $\vec{z}$ direction lies within the optical plane and is parallel to the surface normal.

\begin{equation}
r_{ss} = \frac{n_{0}cos \theta_{0} - n_{1}cos \theta_{1}}{n_{0}cos \theta_{0} + n_{1}cos \theta_{1}}
\label{eq:rss}
\end{equation}

\begin{eqnarray}
r_{pp} = && \frac{n_{1}cos \theta_{0} - n_{0}cos \theta_{1}}
{n_{1}cos \theta_{0} + n_{0}cos \theta_{1}} \nonumber\\
&& + 2iQm_{y}\frac{n_{0}n_{1}cos \theta_{0}sin \theta_{1}} {(n_{1}cos \theta_{0} + n_{0}cos \theta_{1})^2}
\label{eq:rpp}
\end{eqnarray}

\begin{eqnarray}
&&r_{sp} = \nonumber\\
&& iQ\frac{n_{0}n_{1}cos \theta_{0}(m_{x}sin \theta_{1} + m_{z} cos \theta_{1})}{(n_{1}cos \theta_{0} + n_{0}cos \theta_{1})(n_{0}cos \theta_{0} + n_{1}cos \theta_{1})cos \theta_{1}}
\label{eq:rsp}
\end{eqnarray}

\begin{eqnarray}
&&r_{ps} = \nonumber\\
&& -iQ\frac{n_{0}n_{1}cos \theta_{0}(m_{x}sin \theta_{1} - m_{z} cos \theta_{1})}{(n_{1}cos \theta_{0} + n_{0}cos \theta_{1})(n_{0}cos \theta_{0} + n_{1}cos \theta_{1})cos \theta_{1}}
\label{eq:rps}
\end{eqnarray}

$n_{0}$ and $n_{1}$ correspond respectively to the refractive indices of the incident and reflective mediums, $Q$ is the material's complex Voigt constant describing the magneto-optical response, whereas $\theta_{0}$ and $\theta_{1}$ represent respectively the incident and transmitted complex angles.

Jones' formalism \cite{ell_pol_book}, combined with Fresnel's coefficients, allows to derive the final polarization state of a beam passing through the different optical components in any setup, by multiplying each element's associated $(2\times2)$ matrix with the polarization vector in the appropriate order. For a generalized system, the final polarization state given in the $p-s$ basis is determined by expression \ref{eq:normalMOKE}.

\begin{eqnarray}
    \begin{pmatrix}
        E^{f}_{p}\\
        E^{f}_{s}\\
    \end{pmatrix}
    = \text{A}\cdot\underbrace{\text{O}}_{
    \begin{pmatrix}
    r_{pp} & r_{ps} \\
    r_{sp} & r_{ss} \\
    \end{pmatrix}
    }\cdot\vec{\text{b}}
\label{eq:normalMOKE}
\end{eqnarray}

where $\vec{\text{b}}$ is the incident polarization state Jones vector determined by the optical components before reflection, O is the sample's reflection matrix constituted of Fresnel's reflection coefficients, and A denotes the product of matrices for the analyzing optics. The intensity associated to the final electric field vector ($\vec{E}^{f}$) is calculated as given in expression \ref{eq:int}. This quantity has two distinct contributions: non-magnetic and magneto-optical or Kerr signal. The non-magnetic part is solely dependent on the optical properties of the material, \textit{i.e.}, the refractive index, and it is represented by $I_{NM}$. The Kerr contribution is dependent on both the optical and the magneto-optical properties, it can be of first and/or second order with the magnetization (respectively scaling with $Q$ or $Q^2$) depending on the magnetic configuration and the optical elements in the system. Its contribution to the total signal is denoted by $I_{K}$. We restrict our discussions in this work to first order Kerr effect.

\begin{eqnarray}
I = |E^{f}_{p}|^2 + |E^{f}_{s}|^2 = I_{NM} + I_{K}
\label{eq:int}
\end{eqnarray}

\section{Fourier-space resolved MOKE}

\begin{figure}[h]
    \centering
    \includegraphics[scale=0.17]{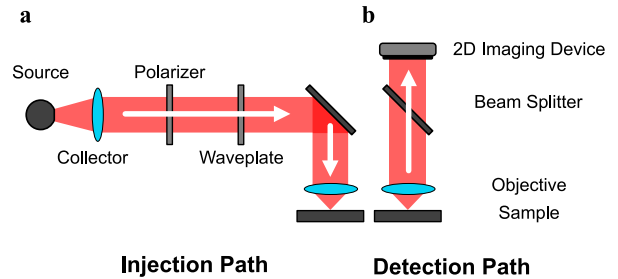}
    \caption{\label{fig:fmoke} Schematics of the Fourier-space resolved Kerr magnetometer diagram, split into (a) sample illumination and (b) light detection path.}
\end{figure}

To perform Fourier MOKE magnetometry we employ a focused beam ellipsometer \cite{ye2007development}, schematically shown in figure \ref{fig:fmoke}. A light source providing a collimated homogeneous intensity profile is used for illumination. The polarization state of the beam is determined by the relative combination of a polarizer and a quarter wave-plate, providing any polarization state ranging from linear to circular, the latter being the one exploited in this work. As in other magneto-optical setups \cite{flajvsman2016high,pathak2014polar}, a high numerical aperture (NA) objective lens is utilized to focus the incoming beam onto the sample's surface down to a diffraction-limited spot size ($s$) given by $s = 1.22\lambda/\text{NA}$, where $\lambda$ is the light's wavelength. The full entrance pupil of the objective is illuminated, providing a wide range of incident $\vec{k}$ on the sample's surface. The reflected beam is collected and collimated by the objective lens before being recorded by a 2D area imaging device with high signal level resolution (\textit{i.e.}, a CCD/CMOS camera). No analyzing optics are included, resulting in a system which provides solely intensity based measurements. Although it may be counter-intuitive at first for those familiar with Kerr magnetometry, this analyzer-free approach will be justified in the following sections.

\begin{figure}[ht]
    \centering
    \includegraphics[scale=0.37]{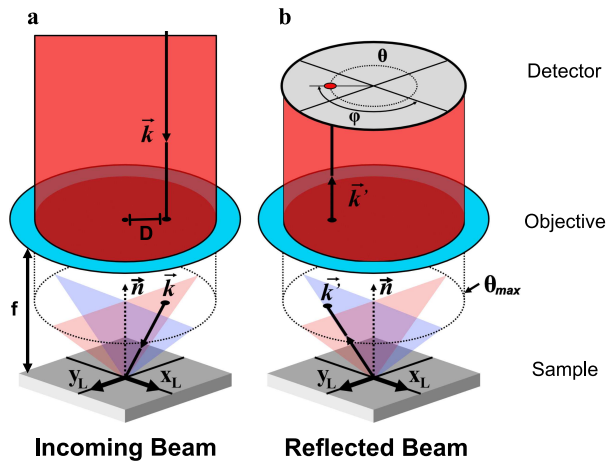}
    \caption{\label{fig:fmoke2} (a) incoming collimated beam reaching the objective lens entrance pupil and focused on the sample's surface. (b) reflected beam collected by the objective lens and reaching the detector, where the optical plane $\varphi$ and incidence angle $\theta$ coordinates are sketched for a single wave-vector $\vec{k}$. Two orthogonal optical planes are shaded in red and blue for illustration purposes.}
\end{figure}

Each pixel in the intensity map recorded on the detector corresponds to a particular incident $\Vec{k}$ on the sample's surface, belonging to a given incidence angle with an associated optical plane. Both these quantities may be described by a 2D coordinate system ($\theta,\varphi$) to map each pixel in the camera to each $\vec{k}$ in a more evident way (figure \ref{fig:fmoke2}). $\theta$, analogous to the polar angle in spherical coordinates, describes the incidence angle on the sample's surface with respect to normal incidence ($0^{\circ}$). The value of $\theta$ can range between $[0^{\circ},\theta_{MAX}]$, where $\theta_{MAX}$ is determined by the NA of the lens. Its value for any pixel is given by $\theta = \sin^{-1}(D/f)$, with $D$ being the distance from the pixel of interest to the centre of the detector, coinciding with the optical axis, and $f$ the focal length of the objective. $\varphi$, analogous to the azimuthal angle in spherical coordinates, describes the projection angle of a given optical plane on the detector with respect to a previously defined in-plane external laboratory reference direction ($x_{L}$). $\varphi$ ranges from $[0^{\circ},360^{\circ}]$, and a single optical plane corresponds to the pair of values $\varphi$ and $\varphi + 180^{\circ}$. 

For the computational modeling of this system, the polarization state, optical elements matrices and magnetization vectors are all expressed in the external laboratory FOR. Although, the light-sample interaction is computed after expressing equations (\ref{eq:rss}-\ref{eq:rps}) and the polarization vector in the FOR of the optical plane, in order to properly compute this interaction. After reflection, the resulting polarization state is re-expressed back into the external laboratory FOR by applying the inverse operation. These changes in FOR are achieved by applying two 2D rotation matrices to the $(x_{L},y_{L})$ components of the vectors, whose rotation angles are respectively $\varphi$ and $-\varphi$. The final polarization vector is thus obtained by including these matrices represented by R($\alpha$) (with $\alpha$ being a rotation angle) into expression \ref{eq:normalMOKE}, resulting in:

\begin{eqnarray}
    \begin{pmatrix}
        E^{f}_{X_{L}}\\
        E^{f}_{Y_{L}}\\
    \end{pmatrix}
    = \text{A}\cdot{\text{R}(-\varphi)}\cdot\text{O}\cdot{\text{R}(+\varphi)}\cdot\vec{\text{b}}
\label{eq:fmoke_fps}
\end{eqnarray}

The intensity for each pixel in the camera $I(\theta,\varphi)$ is thus computed by applying expression \ref{eq:int} to the final polarization state derived in \ref{eq:fmoke_fps}, properly taking into account its associated $(\theta,\varphi)$ values.

\section{Fourier-space resolved intensity maps}

Prior to showing and discussing the procedure for GME, a summary describing the optical and magneto-optical signals generated for given material parameters is shown in this section. This step provides a first look at the intensity profiles recorded by the camera, and its evolution for various scenarios.

\begin{figure*}
\centering
\includegraphics[scale=0.175]{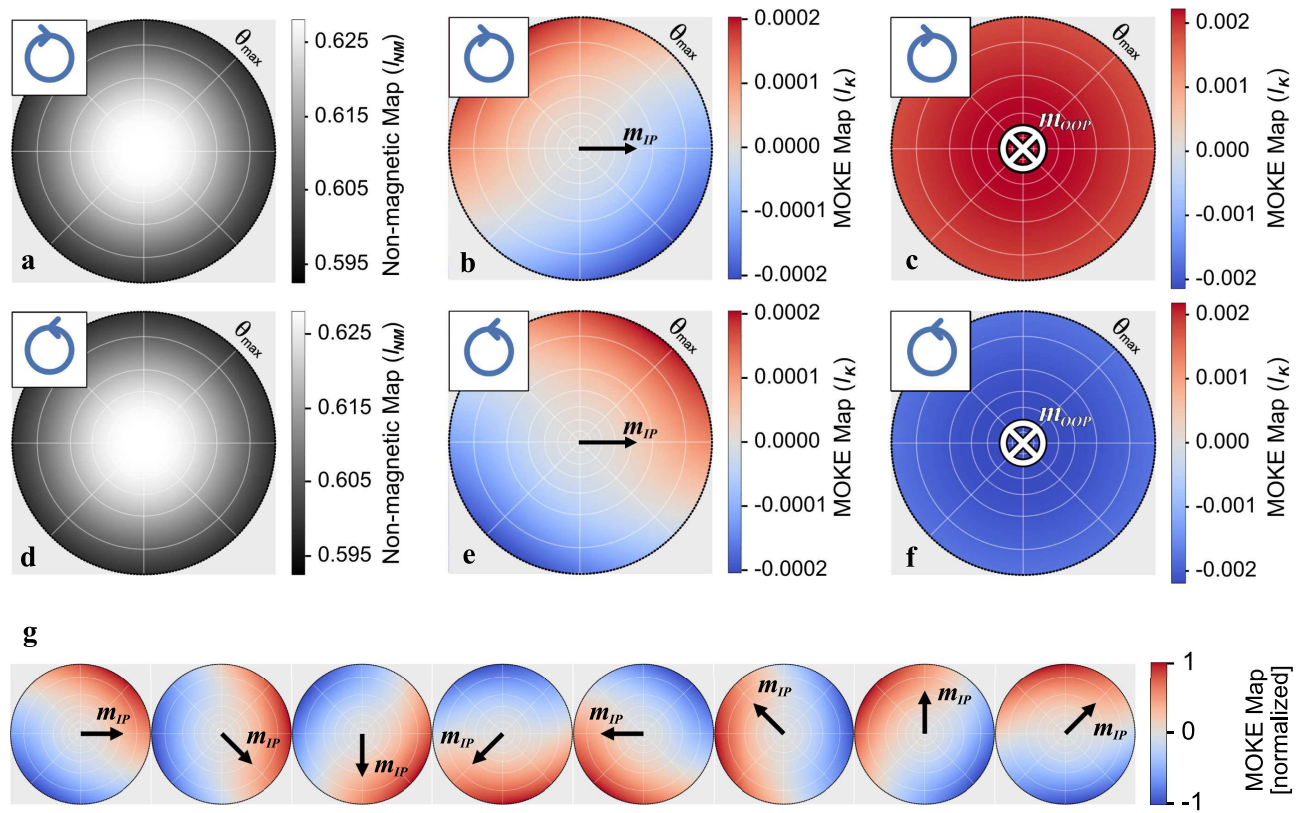}
\caption{\label{fig:circularbasis} (a,d) non-magnetic signal maps for CR and CL incident polarized light without analyzing optics. (b,e) and (c,f) show respectively the magnetic signal maps for CR and CL incident polarization for $\vec{m} = (1,0,0)$ and $\vec{m} = (0,0,1)$ magnetic configurations. (g) evolution of normalized magnetic signal for rotating IP magnetization under CL incident light.}
\end{figure*}

Figure \ref{fig:circularbasis} shows intensity maps for characteristic Permalloy optical and magneto-optical values (800 nm wavelength) \cite{loughran2018enhancing}, for different orientations of the magnetization vector $\vec{m}$ under macrospin approximation, and for both circular eigenmodes, \textit{i.e.}, circularly left (CL) and circularly right (CR). The case of a thick film with thickness significantly greater than the light penetration depth (20 nm \cite{bass2010handbook}) is considered. Each map consists of $201\times201$ pixels, and a value of $\theta_{MAX} = 65^{\circ}$, corresponding to a typical commercial objective NA. The remaining parameters utilized in the simulations are summarized in table \ref{tab:tableCPol}. In all the maps, the central pixel corresponds to normal incidence ($\theta = 0^{\circ}$), and the concentric rings represent pixels with identical $\theta$ values evenly spaced by $10^{\circ}$ intervals. Each radial line corresponds to a different optical plane, given by $\varphi$ and $\varphi + 180^{\circ}$.

\begin{table*}
\begin{ruledtabular}
\begin{tabular}{cccccccccc}
 Map&Map type&Polarization&$n$&$Q$&$m_{X_{L}}$&$m_{Y_{L}}$&$m_{Z_{L}}$\\ 
 \hline
 a & Non-magnetic & CR & 2.25 + i3.7 & 0.006 - i0.011 & 0 & 0 & 0\\
 b & MOKE & CR & \ditto & \ditto & 1 & 0 & 0\\
 c & MOKE & CR & \ditto & \ditto & 0 & 0 & 1\\
 d & Non-magnetic & CL & \ditto & \ditto & 0 & 0 & 0\\
 e & MOKE & CL & \ditto & \ditto & 1 & 0 & 0\\
 f & MOKE & CL & \ditto & \ditto & 0 & 0 & 1\\
\end{tabular}
\end{ruledtabular}
\caption{\label{tab:tableCPol} Summary of simulation parameters for each of the different maps shown in figure \ref{fig:circularbasis}.}
\end{table*}

In the non-magnetic maps, figures \ref{fig:circularbasis} (a,d), the 2D angular intensities $I_{NM}(\theta,\varphi)$ solely depend on $\theta$, but not on $\varphi$. This occurs because circular incident polarization is invariant upon projecting onto different optical planes, \textit{i.e.}, there is no azimuthal symmetry breaking. The non-magnetic intensities satisfy $I_{NM}(\theta,\varphi) = I_{NM}(\theta,\varphi + 180^{\circ})$, and reverting the helicity (represented by the sense of rotation of the circular arrow within the map insets) leaves the maps unaltered. The radial intensity decay is exclusively determined by the material's real and imaginary non-magnetic part of the refractive index via Fresnel's reflection coefficients, yielding as a consequence a unique 2D angular intensity distribution for each material.

The magneto-optical intensity maps, or MOKE maps $I_{K}$, are shown in figures \ref{fig:circularbasis} (b,c,e,f) for Permalloy with either fully IP or OOP magnetic configurations (the non-magnetic contribution is subtracted from the total intensity yielding a purely magneto-optical signal). The total magnetic signal for an arbitrary 3D configuration may be understood as the weighted sum of the intensities for these two orthogonal and independent bases, whose weights depend respectively upon the amount of IP or OOP magnetization. Within this formalism, there is no longer distinction between the traditional IP longitudinal and transverse configurations, and IP magnetization becomes a single configuration on its own.

MOKE maps for a particular IP direction of the magnetization, given by the arrow, are shown for both circular eigenmodes in figures \ref{fig:circularbasis} (b,e). The magnetic intensity 2D angular dependence becomes more complex than in figures \ref{fig:circularbasis} (a,d), due to the azimuthal symmetry breaking of the IP component when projected onto a given optical plane direction. As expected for a system with IP magnetized state, the Kerr signal is maximum at $\theta$ away from normal incidence ($50-60^{\circ}$), and satisfies $I(\theta,\varphi) = - I(\theta,\varphi+180^{\circ})$. The maps show a smooth variation of the intensity pattern as a function of $\varphi$, which in this case has its maximum (in magnitude) at a value of $\varphi$ corresponding to an optical plane in between the ones  parallel and perpendicular to the $\vec{m}$ vector. This particular value of $\varphi$ is exclusively determined by the optical and magneto-optical constants and the magnetization vector. See supplementary material section SM1 for maps generated with other material values, and also a discussion on the effect of these on the value of $\varphi$ which maximizes the IP Kerr signal. Lastly, upon rotation of the IP direction, the magnetic pattern rotates by the same amount as $\vec{m}$ about the normal incidence point, as shown in figure \ref{fig:circularbasis} (g) for CR polarized light.

This behavior contrasts with the one observed for the MOKE maps for the OOP component, as shown in figures \ref{fig:circularbasis} (e,f). These are radially symmetric, \textit{i.e.}, $I_{K}(\theta,\varphi) = I_{K}(\theta,\varphi + 180^{\circ})$, peaking at $\theta = 0^{\circ}$. Unlike for IP components, varying $\varphi$ does not alter the magneto-optical intensity, since the projection of OOP magnetization onto any optical plane direction does not break the azimuthal symmetry. This yields an OOP signal which is only dependent on $\theta$, providing a radial intensity decay. Under helicity reversal, the OOP signal changes sign.

Although linearly polarized light in combination with analyzing optics is more common in MOKE, here, the use of circular light without analyzing optics offers a significant advantage. As just shown, within this approach variations in magnetic intensity from pixel to pixel occur only due to changes in $\theta$ and $\varphi$ for a given $\vec{m}$, unaffected by the incident polarization state. In contrast, linearly polarized light and analyzing optics add an additional azimuthal dependence between the signal and the orientation of the optics involved. This approach would require thus fine tuning of the optical elements in order to maximize magnetic contrast for each $\vec{m}$, making the use of circular incident polarization without analyzing optics more advantageous. Examples of intensity maps under linearly incident polarized light where this effect is observed are shown in the supplementary material section SM2.

\section{Method for Generalized Magneto-Optical Ellipsometry}

In this section the Fourier MOKE technique is exploited to perform GME, by applying a least squares fitting algorithm to the full intensity maps. Utilizing the full maps enable to identify very efficiently the material optical and magneto-optical parameters as well as the 3D magnetization vector, given the strong and unique dependence that exists between these and the 2D angular intensity distribution $I(\theta,\varphi)$. The validity of the procedure is first demonstrated by applying the algorithm to single magnetic state intensity maps, testing the accuracy of the results for different optically and magneto-optically active materials and magnetic configurations. Secondly, this procedure is applied to a magnetic loop where the magnetization evolves over time.

\subsection{GME for a single magnetic state}

The algorithm used here consists of simultaneously fitting the total signal, \textit{i.e.}, non-magnetic plus Kerr signal (expression \ref{eq:int}) for all pixels within the map and for both circular eigenmodes, to the intensity model previously described. Maps for both circular eigenmodes are included in the fitting in order to exploit the helicity-dependent magneto-optical coupling described in the previous section. Additional effects contributing to the total intensity detected in the camera are taken into account by introducing a global offset and a scaling factor. The fit parameters are the normalized magnetization vector components plus the optical and magneto-optical constants. The incident helicity (CL/CR) and the optical plane ($\varphi$) and incidence angle ($\theta$) coordinates, whose values are known at each pixel, serve as independent variables. Refer to supplementary material section SM3 for more details on the algorithm.

\begin{figure*}
\centering
\includegraphics[scale=0.33]{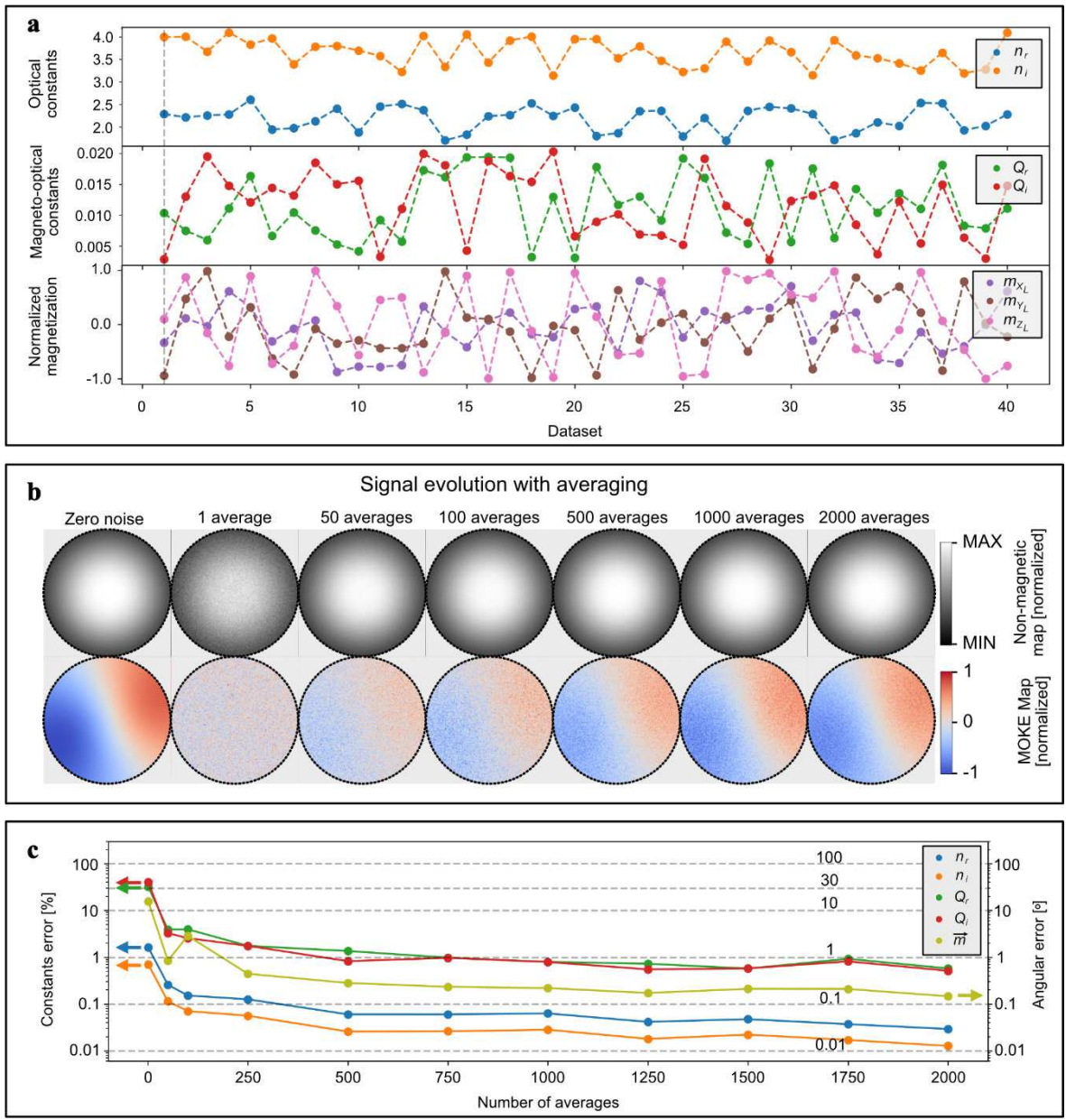}
\caption{\label{fig:fitting_maps_random} (a) values for random optical, magneto-optical and magnetization vector parameters utilized as input for generating the intensity maps. (b) evolution of the optical and magneto-optical signal as a function of the image average number when including sources of noise and discretization. The parameters utilized for these maps correspond to the ones highlighted by the black dashed vertical line in figure (a). (c) average relative error in the fitted parameters for the intensity maps generated with the parameters in figure (a) as a function of the image average number.}
\end{figure*}

In order to test the accuracy of the fitting algorithm in the most general way possible, here it is individually applied to intensity maps generated from a set of 40 different material parameter values, \textit{i.e.}, different optical and magneto-optical constants, as well as single 3D magnetic state configurations. The values for each set of parameters are sampled from a random uniform distribution after setting lower and upper boundaries, in order to preserve their physical meaning. From typical values found in the literature \cite{protopopov2014practical}, these are chosen as $n_{r} = [1.75, 2.75]$, $n_{i} = [3.2, 4.2]$, $Q_{r} = [0.002, 0.02]$, $Q_{i} = [0.002, 0.02]$, with the magnetization vector randomly oriented in 3D space. The value for these randomly generated parameters are displayed in figure \ref{fig:fitting_maps_random} (a). Each of the maps obtained for each dataset has the same size and maximum angle of incidence as the ones shown in the previous section.

The case of zero-noise maps is first considered, \textit{i.e.}, computed directly from the intensity model. When fitting the maps associated to a particular set of material parameters, the fit variables are initialized as completely random before applying the least squares minimizer, as would be done experimentally where no prior knowledge about the sample's magnetic configuration or optical and magneto-optical properties would exist. For further details on the algorithm, refer to supplementary material section SM3. By applying this procedure to each dataset, it is found that the algorithm succeeds in retrieving all the parameter values with precision up to the numerical computer tolerance. An analysis based on numerical methods has been carried out in the supplementary material section SM4, which proves that the best fit provides unique values for these desired parameters.

After showing the validity of the method for a zero-noise case, we now consider some key experimental factors which would make the acquired signals deviate from their ideal values in a real experiment. In particular, the generated maps are modified for a more realistic modeling of the data acquisition process which takes place in a CCD/CMOS camera.

Amongst these sources, signal level resolution is a key factor since the magnetic signal is small in comparison with the dominating large non-magnetic background. For scientific cameras, this is determined by the dynamic range of the device and the analog-to-digital (ADC) converter resolution. The dynamic range is modelled by imposing a finite full well capacity (FWC) for the generated photo-electrons combined with a given read-noise, and the ADC resolution is modelled by registering the intensity values into a total number of discrete levels given in bits, \textit{i.e.}, the number of levels is $2^{bits}$. Photon shot noise due to the discrete nature of photon emission from the light source, as well as the photo-electron generation process described by the quantum-efficiency (QE) have also been modelled, following a similar approach as the one described in python's \textit{Pyxel} package \cite{lucsanyi2018pyxel}. The values for these parameters have been selected based on a typical 16 bit CMOS commercial mid-range scientific camera: $\text{QE} = 61\%$, $\text{FWC} = 40000e^{-}$ and read-noise of $1.5e^{-}$ RMS.

Figure \ref{fig:fitting_maps_random} (b) shows the non-magnetic and MOKE maps for a particular set of parameters, dataset \#1, marked by the black dashed vertical line in figure \ref{fig:fitting_maps_random} (a), as a function of the image average number after having included these sources of noise and discretization (the zero-noise maps are also included for direct comparison). As expected, the single-shot (1 average) images are highly affected by noise, with the magneto-optical intensity map being more affected by Poisson shot noise than the non-magnetic.

An identical fitting procedure as in the zero-noise case is followed for these datasets, obtaining for each file and number of image averages a set of best fit parameters. The relative error for the optical and magneto-optical constants is determined by comparing those obtained from the fitting with their input values. The error in the determination of the magnetization vector is given by the angular separation between input ($\vec{m}_{I}$) and fitted ($\vec{m}_{F}$) vectors.

Figure \ref{fig:fitting_maps_random} (c) shows the average error per parameter for all the files as a function of the image average number. For the single shot, the error obtained in the magnetization vector is around $15^{\circ}$, and $30\%$ for the magneto-optical constants, which contrasts with the under $5\%$ error in the optical constants. This is because the magnetization vector and magneto-optical constants only affect the magneto-optical signal, which is much more influenced by the inclusion of noise than the optical constants, these fully determining the higher signal-to-noise ratio non-magnetic intensity. The error as a function of the image average number shows a decreasing trend, on which the optical parameters maintain a relative error between one and two orders of magnitude smaller than the one corresponding to the magnetization vector and the magneto-optical parameters for this explored range of averages. From approximately 500 averages onwards the error falls below $1\%$ for all parameters for this camera specification.

Other commercial camera characteristic values have been also tested in the fitting algorithm (not shown here), concluding that it is crucial for the camera to have a signal level resolution of at least 16 bits for such great accuracy in determining the magnetic parameters. For lower bit levels, the magneto-optical intensity becomes indistinguishable when noise of this magnitude is added, as a consequence giving a significantly larger relative error.

These results thus demonstrate the validity of this procedure for performing GME without the need of performing reference measurements \cite{vavassori2000polarization}.

\subsection{GME applied to a magnetic loop}

\begin{figure*}
\centering
\includegraphics[scale=0.20]{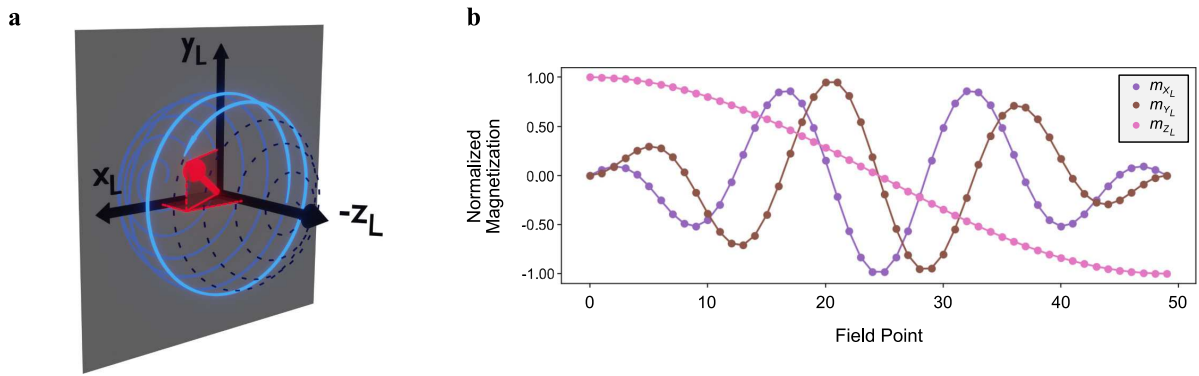}
\caption{\label{fig:finalfig} (a) 3D representation of the magnetization (red arrow) precessing reversal process, and (b) discretization represented in the laboratory FOR.}
\end{figure*}

The algorithm presented in the previous section is now applied to a typical experimental case where the magnetization vector evolves over time. For this, Permalloy characteristic optical and magneto-optical parameters \cite{loughran2018enhancing} are chosen to generate the data (values displayed in table \ref{tab:tableCPol}). The macrospin vector describes the precessing reversal process shown in figure \ref{fig:finalfig} (a), where the magnetization transits from $+m_{z_{L}}$ to $-m_{z_{L}}$ via multiple rotations in a helical fashion through the $(m_{x_{L}},m_{y_{L}})$ plane. The discretization of the input magnetization vector evolution, in 50 field points, is shown as a scatter plot in figure \ref{fig:finalfig} (b). At each field point, intensity maps with the same pixel size and for the same maximum angle of incidence as in the previous section are generated for both circular eigenmodes.

%\begin{figure*}
%\centering
%\includegraphics[scale=0.23]{EPSFigures/GME.eps}
%\caption{\label{fig:finalfig} (a) 3D representation of the magnetization (red arrow) precessing reversal process used to test Fourier MOKE for GME applied to a magnetic loop. (b) representation of fitted (lines) and input (dots) magnetization vector at each field point. (c) Error quantification computed as the magnitude of the difference vector between input and fitted magnetization ($|\vec{m}_{I} - \vec{m}_{F}|$).}
%\end{figure*}

%\begin{figure*}
%\centering
%\includegraphics[scale=0.21]{EPSFigures/GME_opt1.eps}
%\caption{\label{fig:finalfig} (a) 3D representation of the magnetization (red arrow) precessing reversal process, and (b) discretization represented in the laboratory FOR. (c) angular error per field point as a function of the number of image averages.}
%\end{figure*}

For the data analysis, together with the single magnetic state methodology from the previous section, we employ here a self-consistent fitting algorithm which simultaneously takes into account all the field points. For this self-consistent fitting, the fit parameters are the normalized magnetization vector components for each of the field points, plus the optical and magneto-optical constants. Whereas the magnetization is allowed to vary at each field point, the two sets of constants share the same value amongst all the intensity maps, given that these are material constants and thus do not vary.

This procedure is run after including noise and signal discretization, as before, here applied to all intensity maps considered. Further details about the computational implementation of the self-consistent fitting algorithm are given in supplementary material section SM3.

\begin{table*}
\begin{ruledtabular}
\begin{tabular}{ccccccccc}
 \multirow{2}{*}{Parameter}&\multicolumn{2}{c}{1 avg (0.02 s)}&\multicolumn{2}{c}{10 avgs (0.20 s)}&\multicolumn{2}{c}{20 avgs  (0.40 s)}&\multicolumn{2}{c}{250 avgs  (5.00 s)}\\
 \cmidrule(lr){2-3}\cmidrule(lr){4-5}\cmidrule(lr){6-7}\cmidrule(lr){8-9}
 &S&SC&S&SC&S&SC&S&SC\\ \hline
 $n_{r}$ & 1.333$\%$ & 0.137$\%$ & 0.359$\%$ & 0.058$\%$ & 0.189$\%$ & 0.118$\%$ & 0.083$\%$ & 0.003$\%$ \\
 $n_{i}$ & 0.562$\%$ & 0.065$\%$ & 0.148$\%$ & 0.026$\%$ & 0.090$\%$ & 0.048$\%$ & 0.038$\%$ & 0.002$\%$ \\
 $Q_{r}$ & 21.641$\%$ & 2.145$\%$ & 5.867$\%$ & 0.435$\%$ & 3.499$\%$ & 0.231$\%$ & 1.265$\%$ & 0.127$\%$ \\
 $Q_{i}$ & 10.541$\%$ & 0.954$\%$ & 2.842$\%$ & 0.121$\%$&2.034$\%$ & 0.013$\%$ & 0.594$\%$ & 0.042$\%$ \\
 $\vec{m}$ & 6.33$^{\circ}$ & 4.16$^{\circ}$ & 2.52$^{\circ}$ & 1.41$^{\circ}$ & 1.15$^{\circ}$ & 0.94$^{\circ}$ & 0.41$^{\circ}$ & 0.26$^{\circ}$ \\
\end{tabular}
\end{ruledtabular}
\caption{\label{tab:fittable} Median relative errors for optical and magneto-optical constants, and angular error on the magnetization vector obtained for single magnetic state fitting (\textit{S}), and self-consistent approach (\textit{SC}). Next to each average number, the time per field point necessary to acquire that number of averages is shown, assuming a 50 frames per second acquisition time for the scientific camera.}
\end{table*}

The median relative errors for the fitted parameters are displayed for different number of image averages in columns \textit{S} and \textit{SC} of table \ref{tab:fittable}, corresponding respectively to single state and self-consistent fitting procedures. These show that for a given number of averages, the relative error in the self-consistent case is significantly reduced in all parameters in comparison with the single state fitting. The general improvement is the result of simultaneously fitting multiple field points, which effectively acts like averaging of the signals. The self-consistent approach is hence a promising strategy in Fourier MOKE  when performing magnetic loop measurements.

Furthermore, as shown in the previous section, the accuracy on the fitted GME parameters can be further increased by averaging the intensity maps at each field point. Indeed, we observe that a few averages are sufficient to obtain errors in \textit{S} and \textit{SC} comparable to those obtained with conventional GME \cite{berger1997generalized}. The accuracy of the reconstruction of the three dimensional magnetisation vector orientation is also comparable to those obtained with X-ray techniques such as laminography and tomography \cite{donnelly2020time,donnelly2018tomographic,hierro20183d}. These results show Fourier MOKE as a technique with great potential for fast and precise GME experiments. The single-shot (1 average) results also suggest the possibility of expanding GME to dynamic magnetic processes \textit{e.g.} domain wall motion which could include stochastic processes \cite{grollier2020neuromorphic,hayward2015intrinsic}.

\section{Conclusion and Outlook}

In this paper, we propose and computationally investigate the Fourier MOKE technique, based on utilizing a focused beam ellipsometer for spatially resolving the magneto-optical signal in Fourier-space. We investigate the types of optical and magneto-optical maps recorded by a camera using circularly incident polarized light and no analyzing optics for different magnetic states, considering the Kerr effect under linear approximation. By fitting the total intensity maps to this model, we show how we can determine the magnetization vector, as well as the optical and magneto-optical constants with high precision. We envision Fourier MOKE to represent several advantages with respect to conventional GME \cite{kimel20222022}. First, by resolving the Kerr signal in the Fourier space for a wide range of $\vec{k}$, a very accurate fitting of the GME parameters is achieved. Second, our methodology does not require reference measurements (typically taken at saturation). Third, the usage of only two incident polarization states, no modulation optics, and the high accuracy in determining the sample parameters for a small number of image averages would greatly reduce the time of measurement. An added advantage is that our methodology can be implemented not only for a single magnetic state, but for a range of magnetic states which improves the accuracy in the determination of the optical and magneto-optical constants. The ability demonstrated to perform single-shot GME could be of great use for the study of dynamic magnetization effects compatible with camera acquisition rates

The methodology shown here is directly applicable to optically thick single layer films, and could be readily extended to thin and ultra-thin films, as well as to multilayers \cite{qiu1999surface}.

\begin{acknowledgments}
This work was supported by UKRI through an EPSRC studentship, EP/N509668/1 and EP/R513222/1, the European Community under the Horizon 2020 Program, Contract No. 101001290 (3DNANOMAG), the MCIN with funding from European Union NextGenerationEU (PRTR-C17.I1), and the Aragon Government through the Project Q-MAD. Aurelio Hierro-Rodríguez acknowledges the support by Spanish MICIN under grant PID2019-104604RB/AEI/10.13039/501100011033 and by Asturias FICYT under grant AYUD/2021/51185 with the support of FEDER funds. Dédalo Sanz-Hernández acknowledges funding from ANR/CNRS under the French "Plan Relance de l'etat" for the preservation of R\&D. Luka Skoric acknowledges support from the EPSRC Cambridge NanoDTC EP/L015978/1. Charles N. Christensen acknowledges the UK EPSRC Centre for Doctoral Training in Sensor Technologies for a Healthy and Sustainable Future. Claire Donnelly acknowledges funding from the Max Planck Society Lise Meitner Excellence Program.

\end{acknowledgments}

\bibliography{main}% Produces the bibliography via BibTeX.

%apsrev4-2.bst 2019-01-14 (MD) hand-edited version of apsrev4-1.bst
%Control: key (0)
%Control: author (8) initials jnrlst
%Control: editor formatted (1) identically to author
%Control: production of article title (0) allowed
%Control: page (0) single
%Control: year (1) truncated
%Control: production of eprint (0) enabled
\providecommand{\noopsort}[1]{}\providecommand{\singleletter}[1]{#1}%
\begin{thebibliography}{39}%
\makeatletter
\providecommand \@ifxundefined [1]{%
 \@ifx{#1\undefined}
}%
\providecommand \@ifnum [1]{%
 \ifnum #1\expandafter \@firstoftwo
 \else \expandafter \@secondoftwo
 \fi
}%
\providecommand \@ifx [1]{%
 \ifx #1\expandafter \@firstoftwo
 \else \expandafter \@secondoftwo
 \fi
}%
\providecommand \natexlab [1]{#1}%
\providecommand \enquote  [1]{``#1''}%
\providecommand \bibnamefont  [1]{#1}%
\providecommand \bibfnamefont [1]{#1}%
\providecommand \citenamefont [1]{#1}%
\providecommand \href@noop [0]{\@secondoftwo}%
\providecommand \href [0]{\begingroup \@sanitize@url \@href}%
\providecommand \@href[1]{\@@startlink{#1}\@@href}%
\providecommand \@@href[1]{\endgroup#1\@@endlink}%
\providecommand \@sanitize@url [0]{\catcode `\\12\catcode `\$12\catcode
  `\&12\catcode `\#12\catcode `\^12\catcode `\_12\catcode `\%12\relax}%
\providecommand \@@startlink[1]{}%
\providecommand \@@endlink[0]{}%
\providecommand \url  [0]{\begingroup\@sanitize@url \@url }%
\providecommand \@url [1]{\endgroup\@href {#1}{\urlprefix }}%
\providecommand \urlprefix  [0]{URL }%
\providecommand \Eprint [0]{\href }%
\providecommand \doibase [0]{https://doi.org/}%
\providecommand \selectlanguage [0]{\@gobble}%
\providecommand \bibinfo  [0]{\@secondoftwo}%
\providecommand \bibfield  [0]{\@secondoftwo}%
\providecommand \translation [1]{[#1]}%
\providecommand \BibitemOpen [0]{}%
\providecommand \bibitemStop [0]{}%
\providecommand \bibitemNoStop [0]{.\EOS\space}%
\providecommand \EOS [0]{\spacefactor3000\relax}%
\providecommand \BibitemShut  [1]{\csname bibitem#1\endcsname}%
\let\auto@bib@innerbib\@empty
%</preamble>
\bibitem [{\citenamefont {G{\"o}bel}\ \emph {et~al.}(2021)\citenamefont
  {G{\"o}bel}, \citenamefont {Mertig},\ and\ \citenamefont
  {Tretiakov}}]{gobel2021beyond}%
  \BibitemOpen
  \bibfield  {author} {\bibinfo {author} {\bibfnamefont {B.}~\bibnamefont
  {G{\"o}bel}}, \bibinfo {author} {\bibfnamefont {I.}~\bibnamefont {Mertig}},\
  and\ \bibinfo {author} {\bibfnamefont {O.}~\bibnamefont {Tretiakov}},\
  }\bibfield  {title} {\bibinfo {title} {Beyond skyrmions: Review and
  perspectives of alternative magnetic quasiparticles},\ }\href@noop {}
  {\bibfield  {journal} {\bibinfo  {journal} {Physics Reports}\ }\textbf
  {\bibinfo {volume} {895}},\ \bibinfo {pages} {1} (\bibinfo {year}
  {2021})}\BibitemShut {NoStop}%
\bibitem [{\citenamefont {Fern{\'a}ndez-Pacheco}\ \emph
  {et~al.}(2017)\citenamefont {Fern{\'a}ndez-Pacheco}, \citenamefont
  {Streubel}, \citenamefont {Fruchart}, \citenamefont {Hertel}, \citenamefont
  {Fischer},\ and\ \citenamefont {Cowburn}}]{fernandez2017three}%
  \BibitemOpen
  \bibfield  {author} {\bibinfo {author} {\bibfnamefont {A.}~\bibnamefont
  {Fern{\'a}ndez-Pacheco}}, \bibinfo {author} {\bibfnamefont {R.}~\bibnamefont
  {Streubel}}, \bibinfo {author} {\bibfnamefont {O.}~\bibnamefont {Fruchart}},
  \bibinfo {author} {\bibfnamefont {R.}~\bibnamefont {Hertel}}, \bibinfo
  {author} {\bibfnamefont {P.}~\bibnamefont {Fischer}},\ and\ \bibinfo {author}
  {\bibfnamefont {R.}~\bibnamefont {Cowburn}},\ }\bibfield  {title} {\bibinfo
  {title} {Three-dimensional nanomagnetism},\ }\href@noop {} {\bibfield
  {journal} {\bibinfo  {journal} {Nature communications}\ }\textbf {\bibinfo
  {volume} {8}},\ \bibinfo {pages} {1} (\bibinfo {year} {2017})}\BibitemShut
  {NoStop}%
\bibitem [{\citenamefont {Donnelly}\ and\ \citenamefont
  {Scagnoli}(2020)}]{donnelly2020imaging}%
  \BibitemOpen
  \bibfield  {author} {\bibinfo {author} {\bibfnamefont {C.}~\bibnamefont
  {Donnelly}}\ and\ \bibinfo {author} {\bibfnamefont {V.}~\bibnamefont
  {Scagnoli}},\ }\bibfield  {title} {\bibinfo {title} {Imaging
  three-dimensional magnetic systems with x-rays},\ }\href@noop {} {\bibfield
  {journal} {\bibinfo  {journal} {Journal of Physics: Condensed Matter}\
  }\textbf {\bibinfo {volume} {32}},\ \bibinfo {pages} {213001} (\bibinfo
  {year} {2020})}\BibitemShut {NoStop}%
\bibitem [{\citenamefont {McCord}(2015)}]{mccord2015progress}%
  \BibitemOpen
  \bibfield  {author} {\bibinfo {author} {\bibfnamefont {J.}~\bibnamefont
  {McCord}},\ }\bibfield  {title} {\bibinfo {title} {Progress in magnetic
  domain observation by advanced magneto-optical microscopy},\ }\href@noop {}
  {\bibfield  {journal} {\bibinfo  {journal} {Journal of Physics D: Applied
  Physics}\ }\textbf {\bibinfo {volume} {48}},\ \bibinfo {pages} {333001}
  (\bibinfo {year} {2015})}\BibitemShut {NoStop}%
\bibitem [{\citenamefont {Kleemann}(2007)}]{kleemann2007perspective}%
  \BibitemOpen
  \bibfield  {author} {\bibinfo {author} {\bibfnamefont {W.}~\bibnamefont
  {Kleemann}},\ }\bibfield  {title} {\bibinfo {title} {Perspective: Tools of
  modern magnetic materials research: Vector and bragg magneto-optical kerr
  effect for the analysis of nanostructured magnetic arrays [rev. sci. instrum.
  78, 121301 (2007)]},\ }\href@noop {} {\bibfield  {journal} {\bibinfo
  {journal} {Review of Scientific Instruments}\ }\textbf {\bibinfo {volume}
  {78}},\ \bibinfo {pages} {121301} (\bibinfo {year} {2007})}\BibitemShut
  {NoStop}%
\bibitem [{\citenamefont {Teixeira}\ \emph {et~al.}(2011)\citenamefont
  {Teixeira}, \citenamefont {Lusche}, \citenamefont {Ventura}, \citenamefont
  {Fermento}, \citenamefont {Carpinteiro}, \citenamefont {Araujo},
  \citenamefont {Sousa}, \citenamefont {Cardoso},\ and\ \citenamefont
  {Freitas}}]{teixeira2011versatile}%
  \BibitemOpen
  \bibfield  {author} {\bibinfo {author} {\bibfnamefont {J.}~\bibnamefont
  {Teixeira}}, \bibinfo {author} {\bibfnamefont {R.}~\bibnamefont {Lusche}},
  \bibinfo {author} {\bibfnamefont {J.}~\bibnamefont {Ventura}}, \bibinfo
  {author} {\bibfnamefont {R.}~\bibnamefont {Fermento}}, \bibinfo {author}
  {\bibfnamefont {F.}~\bibnamefont {Carpinteiro}}, \bibinfo {author}
  {\bibfnamefont {J.}~\bibnamefont {Araujo}}, \bibinfo {author} {\bibfnamefont
  {J.}~\bibnamefont {Sousa}}, \bibinfo {author} {\bibfnamefont
  {S.}~\bibnamefont {Cardoso}},\ and\ \bibinfo {author} {\bibfnamefont
  {P.}~\bibnamefont {Freitas}},\ }\bibfield  {title} {\bibinfo {title}
  {Versatile, high sensitivity, and automatized angular dependent vectorial
  kerr magnetometer for the analysis of nanostructured materials},\ }\href@noop
  {} {\bibfield  {journal} {\bibinfo  {journal} {Review of Scientific
  Instruments}\ }\textbf {\bibinfo {volume} {82}},\ \bibinfo {pages} {043902}
  (\bibinfo {year} {2011})}\BibitemShut {NoStop}%
\bibitem [{\citenamefont {Jim{\'e}nez}\ \emph {et~al.}(2014)\citenamefont
  {Jim{\'e}nez}, \citenamefont {Mikuszeit}, \citenamefont {Cu{\~n}ado},
  \citenamefont {Perna}, \citenamefont {Pedrosa}, \citenamefont {Maccariello},
  \citenamefont {Rodrigo}, \citenamefont {Ni{\~n}o}, \citenamefont {Bollero},
  \citenamefont {Camarero} \emph {et~al.}}]{jimenez2014vectorial}%
  \BibitemOpen
  \bibfield  {author} {\bibinfo {author} {\bibfnamefont {E.}~\bibnamefont
  {Jim{\'e}nez}}, \bibinfo {author} {\bibfnamefont {N.}~\bibnamefont
  {Mikuszeit}}, \bibinfo {author} {\bibfnamefont {J.}~\bibnamefont
  {Cu{\~n}ado}}, \bibinfo {author} {\bibfnamefont {P.}~\bibnamefont {Perna}},
  \bibinfo {author} {\bibfnamefont {J.}~\bibnamefont {Pedrosa}}, \bibinfo
  {author} {\bibfnamefont {D.}~\bibnamefont {Maccariello}}, \bibinfo {author}
  {\bibfnamefont {C.}~\bibnamefont {Rodrigo}}, \bibinfo {author} {\bibfnamefont
  {M.}~\bibnamefont {Ni{\~n}o}}, \bibinfo {author} {\bibfnamefont
  {A.}~\bibnamefont {Bollero}}, \bibinfo {author} {\bibfnamefont
  {J.}~\bibnamefont {Camarero}}, \emph {et~al.},\ }\bibfield  {title} {\bibinfo
  {title} {Vectorial kerr magnetometer for simultaneous and quantitative
  measurements of the in-plane magnetization components},\ }\href@noop {}
  {\bibfield  {journal} {\bibinfo  {journal} {Review of Scientific
  Instruments}\ }\textbf {\bibinfo {volume} {85}},\ \bibinfo {pages} {053904}
  (\bibinfo {year} {2014})}\BibitemShut {NoStop}%
\bibitem [{\citenamefont {Allwood}\ \emph {et~al.}(2003)\citenamefont
  {Allwood}, \citenamefont {Xiong}, \citenamefont {Cooke},\ and\ \citenamefont
  {Cowburn}}]{allwood2003magneto}%
  \BibitemOpen
  \bibfield  {author} {\bibinfo {author} {\bibfnamefont {D.}~\bibnamefont
  {Allwood}}, \bibinfo {author} {\bibfnamefont {G.}~\bibnamefont {Xiong}},
  \bibinfo {author} {\bibfnamefont {M.}~\bibnamefont {Cooke}},\ and\ \bibinfo
  {author} {\bibfnamefont {R.}~\bibnamefont {Cowburn}},\ }\bibfield  {title}
  {\bibinfo {title} {Magneto-optical kerr effect analysis of magnetic
  nanostructures},\ }\href@noop {} {\bibfield  {journal} {\bibinfo  {journal}
  {Journal of Physics D: Applied Physics}\ }\textbf {\bibinfo {volume} {36}},\
  \bibinfo {pages} {2175} (\bibinfo {year} {2003})}\BibitemShut {NoStop}%
\bibitem [{\citenamefont {Flaj{\v{s}}man}\ \emph {et~al.}(2016)\citenamefont
  {Flaj{\v{s}}man}, \citenamefont {Urb{\'a}nek}, \citenamefont
  {K{\v{r}}i{\v{z}}{\'a}kov{\'a}}, \citenamefont {Va{\v{n}}atka}, \citenamefont
  {Tur{\v{c}}an},\ and\ \citenamefont {{\v{S}}ikola}}]{flajvsman2016high}%
  \BibitemOpen
  \bibfield  {author} {\bibinfo {author} {\bibfnamefont {L.}~\bibnamefont
  {Flaj{\v{s}}man}}, \bibinfo {author} {\bibfnamefont {M.}~\bibnamefont
  {Urb{\'a}nek}}, \bibinfo {author} {\bibfnamefont {V.}~\bibnamefont
  {K{\v{r}}i{\v{z}}{\'a}kov{\'a}}}, \bibinfo {author} {\bibfnamefont
  {M.}~\bibnamefont {Va{\v{n}}atka}}, \bibinfo {author} {\bibfnamefont
  {I.}~\bibnamefont {Tur{\v{c}}an}},\ and\ \bibinfo {author} {\bibfnamefont
  {T.}~\bibnamefont {{\v{S}}ikola}},\ }\bibfield  {title} {\bibinfo {title}
  {High-resolution fully vectorial scanning kerr magnetometer},\ }\href@noop {}
  {\bibfield  {journal} {\bibinfo  {journal} {Review of Scientific
  Instruments}\ }\textbf {\bibinfo {volume} {87}},\ \bibinfo {pages} {053704}
  (\bibinfo {year} {2016})}\BibitemShut {NoStop}%
\bibitem [{\citenamefont {Soldatov}\ and\ \citenamefont
  {Sch{\"a}fer}(2017{\natexlab{a}})}]{soldatov2017selective}%
  \BibitemOpen
  \bibfield  {author} {\bibinfo {author} {\bibfnamefont {I.}~\bibnamefont
  {Soldatov}}\ and\ \bibinfo {author} {\bibfnamefont {R.}~\bibnamefont
  {Sch{\"a}fer}},\ }\bibfield  {title} {\bibinfo {title} {Selective sensitivity
  in kerr microscopy},\ }\href@noop {} {\bibfield  {journal} {\bibinfo
  {journal} {Review of Scientific Instruments}\ }\textbf {\bibinfo {volume}
  {88}},\ \bibinfo {pages} {073701} (\bibinfo {year}
  {2017}{\natexlab{a}})}\BibitemShut {NoStop}%
\bibitem [{\citenamefont {Choudhary}\ \emph {et~al.}(2022)\citenamefont
  {Choudhary}, \citenamefont {Jansche}, \citenamefont {Grubesa}, \citenamefont
  {Trier}, \citenamefont {Goll}, \citenamefont {Bernthaler},\ and\
  \citenamefont {Schneider}}]{choudhary2022grain}%
  \BibitemOpen
  \bibfield  {author} {\bibinfo {author} {\bibfnamefont {A.}~\bibnamefont
  {Choudhary}}, \bibinfo {author} {\bibfnamefont {A.}~\bibnamefont {Jansche}},
  \bibinfo {author} {\bibfnamefont {T.}~\bibnamefont {Grubesa}}, \bibinfo
  {author} {\bibfnamefont {F.}~\bibnamefont {Trier}}, \bibinfo {author}
  {\bibfnamefont {D.}~\bibnamefont {Goll}}, \bibinfo {author} {\bibfnamefont
  {T.}~\bibnamefont {Bernthaler}},\ and\ \bibinfo {author} {\bibfnamefont
  {G.}~\bibnamefont {Schneider}},\ }\bibfield  {title} {\bibinfo {title} {Grain
  size analysis in permanent magnets from kerr microscopy images using machine
  learning techniques},\ }\href@noop {} {\bibfield  {journal} {\bibinfo
  {journal} {Materials Characterization}\ ,\ \bibinfo {pages} {111790}}
  (\bibinfo {year} {2022})}\BibitemShut {NoStop}%
\bibitem [{\citenamefont {Soldatov}\ and\ \citenamefont
  {Sch{\"a}fer}(2017{\natexlab{b}})}]{soldatov2017advanced}%
  \BibitemOpen
  \bibfield  {author} {\bibinfo {author} {\bibfnamefont {I.}~\bibnamefont
  {Soldatov}}\ and\ \bibinfo {author} {\bibfnamefont {R.}~\bibnamefont
  {Sch{\"a}fer}},\ }\bibfield  {title} {\bibinfo {title} {Advanced moke
  magnetometry in wide-field kerr-microscopy},\ }\href@noop {} {\bibfield
  {journal} {\bibinfo  {journal} {Journal of Applied Physics}\ }\textbf
  {\bibinfo {volume} {122}},\ \bibinfo {pages} {153906} (\bibinfo {year}
  {2017}{\natexlab{b}})}\BibitemShut {NoStop}%
\bibitem [{\citenamefont {Idigoras}\ \emph {et~al.}(2010)\citenamefont
  {Idigoras}, \citenamefont {Vavassori}, \citenamefont {Porro},\ and\
  \citenamefont {Berger}}]{idigoras2010kerr}%
  \BibitemOpen
  \bibfield  {author} {\bibinfo {author} {\bibfnamefont {O.}~\bibnamefont
  {Idigoras}}, \bibinfo {author} {\bibfnamefont {P.}~\bibnamefont {Vavassori}},
  \bibinfo {author} {\bibfnamefont {J.}~\bibnamefont {Porro}},\ and\ \bibinfo
  {author} {\bibfnamefont {A.}~\bibnamefont {Berger}},\ }\bibfield  {title}
  {\bibinfo {title} {Kerr microscopy study of magnetization reversal in
  uniaxial co-films},\ }\href@noop {} {\bibfield  {journal} {\bibinfo
  {journal} {Journal of magnetism and magnetic materials}\ }\textbf {\bibinfo
  {volume} {322}},\ \bibinfo {pages} {L57} (\bibinfo {year}
  {2010})}\BibitemShut {NoStop}%
\bibitem [{\citenamefont {Ding}\ \emph {et~al.}(2000)\citenamefont {Ding},
  \citenamefont {P{\"u}tter}, \citenamefont {Oepen},\ and\ \citenamefont
  {Kirschner}}]{ding2000experimental}%
  \BibitemOpen
  \bibfield  {author} {\bibinfo {author} {\bibfnamefont {H.}~\bibnamefont
  {Ding}}, \bibinfo {author} {\bibfnamefont {S.}~\bibnamefont {P{\"u}tter}},
  \bibinfo {author} {\bibfnamefont {H.}~\bibnamefont {Oepen}},\ and\ \bibinfo
  {author} {\bibfnamefont {J.}~\bibnamefont {Kirschner}},\ }\bibfield  {title}
  {\bibinfo {title} {Experimental method for separating longitudinal and polar
  kerr signals},\ }\href@noop {} {\bibfield  {journal} {\bibinfo  {journal}
  {Journal of magnetism and magnetic materials}\ }\textbf {\bibinfo {volume}
  {212}},\ \bibinfo {pages} {5} (\bibinfo {year} {2000})}\BibitemShut {NoStop}%
\bibitem [{\citenamefont {Vavassori}(2000)}]{vavassori2000polarization}%
  \BibitemOpen
  \bibfield  {author} {\bibinfo {author} {\bibfnamefont {P.}~\bibnamefont
  {Vavassori}},\ }\bibfield  {title} {\bibinfo {title} {Polarization modulation
  technique for magneto-optical quantitative vector magnetometry},\ }\href@noop
  {} {\bibfield  {journal} {\bibinfo  {journal} {Applied Physics Letters}\
  }\textbf {\bibinfo {volume} {77}},\ \bibinfo {pages} {1605} (\bibinfo {year}
  {2000})}\BibitemShut {NoStop}%
\bibitem [{\citenamefont {Berger}\ and\ \citenamefont
  {Pufall}(1997)}]{berger1997generalized}%
  \BibitemOpen
  \bibfield  {author} {\bibinfo {author} {\bibfnamefont {A.}~\bibnamefont
  {Berger}}\ and\ \bibinfo {author} {\bibfnamefont {M.}~\bibnamefont
  {Pufall}},\ }\bibfield  {title} {\bibinfo {title} {Generalized
  magneto-optical ellipsometry},\ }\href@noop {} {\bibfield  {journal}
  {\bibinfo  {journal} {Applied physics letters}\ }\textbf {\bibinfo {volume}
  {71}},\ \bibinfo {pages} {965} (\bibinfo {year} {1997})}\BibitemShut
  {NoStop}%
\bibitem [{\citenamefont {Oblak}\ \emph {et~al.}(2020)\citenamefont {Oblak},
  \citenamefont {Riego}, \citenamefont {Garcia-Manso}, \citenamefont
  {Mart{\'\i}nez-de Guerenu}, \citenamefont {Arizti}, \citenamefont {Artetxe},\
  and\ \citenamefont {Berger}}]{oblak2020ultrasensitive}%
  \BibitemOpen
  \bibfield  {author} {\bibinfo {author} {\bibfnamefont {E.}~\bibnamefont
  {Oblak}}, \bibinfo {author} {\bibfnamefont {P.}~\bibnamefont {Riego}},
  \bibinfo {author} {\bibfnamefont {A.}~\bibnamefont {Garcia-Manso}}, \bibinfo
  {author} {\bibfnamefont {A.}~\bibnamefont {Mart{\'\i}nez-de Guerenu}},
  \bibinfo {author} {\bibfnamefont {F.}~\bibnamefont {Arizti}}, \bibinfo
  {author} {\bibfnamefont {I.}~\bibnamefont {Artetxe}},\ and\ \bibinfo {author}
  {\bibfnamefont {A.}~\bibnamefont {Berger}},\ }\bibfield  {title} {\bibinfo
  {title} {Ultrasensitive transverse magneto-optical kerr effect measurements
  using an effective ellipsometric detection scheme},\ }\href@noop {}
  {\bibfield  {journal} {\bibinfo  {journal} {Journal of Physics D: Applied
  Physics}\ }\textbf {\bibinfo {volume} {53}},\ \bibinfo {pages} {205001}
  (\bibinfo {year} {2020})}\BibitemShut {NoStop}%
\bibitem [{\citenamefont {Neuber}\ \emph {et~al.}(2003)\citenamefont {Neuber},
  \citenamefont {Rauer}, \citenamefont {Kunze}, \citenamefont {Korn},
  \citenamefont {Pels}, \citenamefont {Meier}, \citenamefont {Merkt},
  \citenamefont {B{\"a}ckstr{\"o}m},\ and\ \citenamefont
  {R{\"u}bhausen}}]{neuber2003temperature}%
  \BibitemOpen
  \bibfield  {author} {\bibinfo {author} {\bibfnamefont {G.}~\bibnamefont
  {Neuber}}, \bibinfo {author} {\bibfnamefont {R.}~\bibnamefont {Rauer}},
  \bibinfo {author} {\bibfnamefont {J.}~\bibnamefont {Kunze}}, \bibinfo
  {author} {\bibfnamefont {T.}~\bibnamefont {Korn}}, \bibinfo {author}
  {\bibfnamefont {C.}~\bibnamefont {Pels}}, \bibinfo {author} {\bibfnamefont
  {G.}~\bibnamefont {Meier}}, \bibinfo {author} {\bibfnamefont
  {U.}~\bibnamefont {Merkt}}, \bibinfo {author} {\bibfnamefont
  {J.}~\bibnamefont {B{\"a}ckstr{\"o}m}},\ and\ \bibinfo {author}
  {\bibfnamefont {M.}~\bibnamefont {R{\"u}bhausen}},\ }\bibfield  {title}
  {\bibinfo {title} {Temperature-dependent spectral generalized magneto-optical
  ellipsometry},\ }\href@noop {} {\bibfield  {journal} {\bibinfo  {journal}
  {Applied Physics Letters}\ }\textbf {\bibinfo {volume} {83}},\ \bibinfo
  {pages} {4509} (\bibinfo {year} {2003})}\BibitemShut {NoStop}%
\bibitem [{\citenamefont {Mok}\ \emph {et~al.}(2011)\citenamefont {Mok},
  \citenamefont {Du},\ and\ \citenamefont {Schmidt}}]{mok2011vector}%
  \BibitemOpen
  \bibfield  {author} {\bibinfo {author} {\bibfnamefont {K.}~\bibnamefont
  {Mok}}, \bibinfo {author} {\bibfnamefont {N.}~\bibnamefont {Du}},\ and\
  \bibinfo {author} {\bibfnamefont {H.}~\bibnamefont {Schmidt}},\ }\bibfield
  {title} {\bibinfo {title} {Vector-magneto-optical generalized ellipsometry},\
  }\href@noop {} {\bibfield  {journal} {\bibinfo  {journal} {Review of
  Scientific Instruments}\ }\textbf {\bibinfo {volume} {82}},\ \bibinfo {pages}
  {033112} (\bibinfo {year} {2011})}\BibitemShut {NoStop}%
\bibitem [{\citenamefont {Arregi}\ \emph {et~al.}(2015)\citenamefont {Arregi},
  \citenamefont {Gonz{\'a}lez-D{\'\i}az}, \citenamefont {Idigoras},\ and\
  \citenamefont {Berger}}]{arregi2015strain}%
  \BibitemOpen
  \bibfield  {author} {\bibinfo {author} {\bibfnamefont {J.}~\bibnamefont
  {Arregi}}, \bibinfo {author} {\bibfnamefont {J.}~\bibnamefont
  {Gonz{\'a}lez-D{\'\i}az}}, \bibinfo {author} {\bibfnamefont {O.}~\bibnamefont
  {Idigoras}},\ and\ \bibinfo {author} {\bibfnamefont {A.}~\bibnamefont
  {Berger}},\ }\bibfield  {title} {\bibinfo {title} {Strain-induced
  magneto-optical anisotropy in epitaxial hcp co films},\ }\href@noop {}
  {\bibfield  {journal} {\bibinfo  {journal} {Physical Review B}\ }\textbf
  {\bibinfo {volume} {92}},\ \bibinfo {pages} {184405} (\bibinfo {year}
  {2015})}\BibitemShut {NoStop}%
\bibitem [{\citenamefont {Berger}\ and\ \citenamefont
  {Pufall}(1999)}]{berger1999quantitative}%
  \BibitemOpen
  \bibfield  {author} {\bibinfo {author} {\bibfnamefont {A.}~\bibnamefont
  {Berger}}\ and\ \bibinfo {author} {\bibfnamefont {M.}~\bibnamefont
  {Pufall}},\ }\bibfield  {title} {\bibinfo {title} {Quantitative vector
  magnetometry using generalized magneto-optical ellipsometry},\ }\href@noop {}
  {\bibfield  {journal} {\bibinfo  {journal} {Journal of applied physics}\
  }\textbf {\bibinfo {volume} {85}},\ \bibinfo {pages} {4583} (\bibinfo {year}
  {1999})}\BibitemShut {NoStop}%
\bibitem [{\citenamefont {Ye}\ \emph {et~al.}(2007)\citenamefont {Ye},
  \citenamefont {Kwak}, \citenamefont {Kim}, \citenamefont {Cho}, \citenamefont
  {Cho},\ and\ \citenamefont {Chegal}}]{ye2007development}%
  \BibitemOpen
  \bibfield  {author} {\bibinfo {author} {\bibfnamefont {S.}~\bibnamefont
  {Ye}}, \bibinfo {author} {\bibfnamefont {Y.}~\bibnamefont {Kwak}}, \bibinfo
  {author} {\bibfnamefont {S.}~\bibnamefont {Kim}}, \bibinfo {author}
  {\bibfnamefont {H.}~\bibnamefont {Cho}}, \bibinfo {author} {\bibfnamefont
  {Y.}~\bibnamefont {Cho}},\ and\ \bibinfo {author} {\bibfnamefont
  {W.}~\bibnamefont {Chegal}},\ }\bibfield  {title} {\bibinfo {title}
  {Development of a focused-beam ellipsometer based on a new principle},\ }in\
  \href@noop {} {\emph {\bibinfo {booktitle} {AIP Conference Proceedings}}},\
  Vol.\ \bibinfo {volume} {931}\ (\bibinfo {organization} {American Institute
  of Physics},\ \bibinfo {year} {2007})\ pp.\ \bibinfo {pages}
  {69--73}\BibitemShut {NoStop}%
\bibitem [{\citenamefont {Ye}\ \emph {et~al.}(2008)\citenamefont {Ye},
  \citenamefont {Kim}, \citenamefont {Kwak}, \citenamefont {Cho}, \citenamefont
  {Cho},\ and\ \citenamefont {Chegal}}]{ye2008ellipsometric}%
  \BibitemOpen
  \bibfield  {author} {\bibinfo {author} {\bibfnamefont {S.}~\bibnamefont
  {Ye}}, \bibinfo {author} {\bibfnamefont {S.}~\bibnamefont {Kim}}, \bibinfo
  {author} {\bibfnamefont {Y.}~\bibnamefont {Kwak}}, \bibinfo {author}
  {\bibfnamefont {H.}~\bibnamefont {Cho}}, \bibinfo {author} {\bibfnamefont
  {Y.}~\bibnamefont {Cho}},\ and\ \bibinfo {author} {\bibfnamefont
  {W.}~\bibnamefont {Chegal}},\ }\bibfield  {title} {\bibinfo {title} {An
  ellipsometric data acquisition method for thin film thickness measurement in
  real time},\ }\href@noop {} {\bibfield  {journal} {\bibinfo  {journal}
  {Measurement Science and Technology}\ }\textbf {\bibinfo {volume} {19}},\
  \bibinfo {pages} {047002} (\bibinfo {year} {2008})}\BibitemShut {NoStop}%
\bibitem [{\citenamefont {Lee}\ \emph {et~al.}(2020)\citenamefont {Lee},
  \citenamefont {Lee}, \citenamefont {Choi},\ and\ \citenamefont
  {Pahk}}]{lee2020co}%
  \BibitemOpen
  \bibfield  {author} {\bibinfo {author} {\bibfnamefont {S.}~\bibnamefont
  {Lee}}, \bibinfo {author} {\bibfnamefont {S.}~\bibnamefont {Lee}}, \bibinfo
  {author} {\bibfnamefont {G.}~\bibnamefont {Choi}},\ and\ \bibinfo {author}
  {\bibfnamefont {H.}~\bibnamefont {Pahk}},\ }\bibfield  {title} {\bibinfo
  {title} {Co-axial spectroscopic snap-shot ellipsometry for real-time
  thickness measurements with a small spot size},\ }\href@noop {} {\bibfield
  {journal} {\bibinfo  {journal} {Optics Express}\ }\textbf {\bibinfo {volume}
  {28}},\ \bibinfo {pages} {25879} (\bibinfo {year} {2020})}\BibitemShut
  {NoStop}%
\bibitem [{\citenamefont {Protopopov}(2014)}]{protopopov2014practical}%
  \BibitemOpen
  \bibfield  {author} {\bibinfo {author} {\bibfnamefont {V.}~\bibnamefont
  {Protopopov}},\ }\bibinfo {title} {Practical opto-electronics}\ (\bibinfo
  {publisher} {Springer},\ \bibinfo {year} {2014})\ pp.\ \bibinfo {pages}
  {232--235}\BibitemShut {NoStop}%
\bibitem [{\citenamefont {Qiu}\ and\ \citenamefont
  {Bader}(1999)}]{qiu1999surface}%
  \BibitemOpen
  \bibfield  {author} {\bibinfo {author} {\bibfnamefont {Z.}~\bibnamefont
  {Qiu}}\ and\ \bibinfo {author} {\bibfnamefont {S.~D.}\ \bibnamefont
  {Bader}},\ }\bibfield  {title} {\bibinfo {title} {Surface magneto-optic kerr
  effect (smoke)},\ }\href@noop {} {\bibfield  {journal} {\bibinfo  {journal}
  {Journal of magnetism and magnetic materials}\ }\textbf {\bibinfo {volume}
  {200}},\ \bibinfo {pages} {664} (\bibinfo {year} {1999})}\BibitemShut
  {NoStop}%
\bibitem [{\citenamefont {Wu}\ \emph {et~al.}(2000)\citenamefont {Wu},
  \citenamefont {Moore},\ and\ \citenamefont {Hicken}}]{wu2000optical}%
  \BibitemOpen
  \bibfield  {author} {\bibinfo {author} {\bibfnamefont {J.}~\bibnamefont
  {Wu}}, \bibinfo {author} {\bibfnamefont {J.}~\bibnamefont {Moore}},\ and\
  \bibinfo {author} {\bibfnamefont {R.}~\bibnamefont {Hicken}},\ }\bibfield
  {title} {\bibinfo {title} {Optical pump-probe studies of the rise and damping
  of ferromagnetic resonance oscillations in a thin fe film},\ }\href@noop {}
  {\bibfield  {journal} {\bibinfo  {journal} {Journal of magnetism and magnetic
  materials}\ }\textbf {\bibinfo {volume} {222}},\ \bibinfo {pages} {189}
  (\bibinfo {year} {2000})}\BibitemShut {NoStop}%
\bibitem [{\citenamefont {Loughran}\ \emph {et~al.}(2018)\citenamefont
  {Loughran}, \citenamefont {Keatley}, \citenamefont {Hendry}, \citenamefont
  {Barnes},\ and\ \citenamefont {Hicken}}]{loughran2018enhancing}%
  \BibitemOpen
  \bibfield  {author} {\bibinfo {author} {\bibfnamefont {T.}~\bibnamefont
  {Loughran}}, \bibinfo {author} {\bibfnamefont {P.}~\bibnamefont {Keatley}},
  \bibinfo {author} {\bibfnamefont {E.}~\bibnamefont {Hendry}}, \bibinfo
  {author} {\bibfnamefont {W.}~\bibnamefont {Barnes}},\ and\ \bibinfo {author}
  {\bibfnamefont {R.}~\bibnamefont {Hicken}},\ }\bibfield  {title} {\bibinfo
  {title} {Enhancing the magneto-optical kerr effect through the use of a
  plasmonic antenna},\ }\href@noop {} {\bibfield  {journal} {\bibinfo
  {journal} {Optics express}\ }\textbf {\bibinfo {volume} {26}},\ \bibinfo
  {pages} {4738} (\bibinfo {year} {2018})}\BibitemShut {NoStop}%
\bibitem [{\citenamefont {Hubert}\ and\ \citenamefont
  {Sch{\"a}fer}(2008)}]{hubert2008magnetic}%
  \BibitemOpen
  \bibfield  {author} {\bibinfo {author} {\bibfnamefont {A.}~\bibnamefont
  {Hubert}}\ and\ \bibinfo {author} {\bibfnamefont {R.}~\bibnamefont
  {Sch{\"a}fer}},\ }\href@noop {} {\emph {\bibinfo {title} {Magnetic domains:
  the analysis of magnetic microstructures}}}\ (\bibinfo  {publisher} {Springer
  Science \& Business Media},\ \bibinfo {year} {2008})\ pp.\ \bibinfo {pages}
  {24--25}\BibitemShut {NoStop}%
\bibitem [{\citenamefont {Azzam}\ and\ \citenamefont
  {Bashara}(1977)}]{ell_pol_book}%
  \BibitemOpen
  \bibfield  {author} {\bibinfo {author} {\bibfnamefont {R.~M.~A.}\
  \bibnamefont {Azzam}}\ and\ \bibinfo {author} {\bibfnamefont {N.~M.}\
  \bibnamefont {Bashara}},\ }\href@noop {} {\emph {\bibinfo {title}
  {Ellipsometry and polarized light}}}\ (\bibinfo  {publisher} {North-Holland
  Pub. Co.},\ \bibinfo {year} {1977})\BibitemShut {NoStop}%
\bibitem [{\citenamefont {Pathak}\ and\ \citenamefont
  {Sharma}(2014)}]{pathak2014polar}%
  \BibitemOpen
  \bibfield  {author} {\bibinfo {author} {\bibfnamefont {S.}~\bibnamefont
  {Pathak}}\ and\ \bibinfo {author} {\bibfnamefont {M.}~\bibnamefont
  {Sharma}},\ }\bibfield  {title} {\bibinfo {title} {Polar magneto-optical kerr
  effect instrument for 1-dimensional magnetic nanostructures},\ }\href@noop {}
  {\bibfield  {journal} {\bibinfo  {journal} {Journal of Applied Physics}\
  }\textbf {\bibinfo {volume} {115}},\ \bibinfo {pages} {043906} (\bibinfo
  {year} {2014})}\BibitemShut {NoStop}%
\bibitem [{\citenamefont {Bass}(2010)}]{bass2010handbook}%
  \BibitemOpen
  \bibfield  {author} {\bibinfo {author} {\bibfnamefont {M.}~\bibnamefont
  {Bass}},\ }\href@noop {} {\emph {\bibinfo {title} {Handbook of optics: volume
  ii-design, fabrication, and testing; sources and detectors; radiometry and
  photometry}}}\ (\bibinfo  {publisher} {McGraw-Hill Education},\ \bibinfo
  {year} {2010})\BibitemShut {NoStop}%
\bibitem [{\citenamefont {Lucsanyi}\ \emph {et~al.}(2018)\citenamefont
  {Lucsanyi}, \citenamefont {Prod'homme}, \citenamefont {Smit}, \citenamefont
  {Lemmel}, \citenamefont {Crouzet}, \citenamefont {Verhoeve},\ and\
  \citenamefont {Shortt}}]{lucsanyi2018pyxel}%
  \BibitemOpen
  \bibfield  {author} {\bibinfo {author} {\bibfnamefont {D.}~\bibnamefont
  {Lucsanyi}}, \bibinfo {author} {\bibfnamefont {T.}~\bibnamefont
  {Prod'homme}}, \bibinfo {author} {\bibfnamefont {H.}~\bibnamefont {Smit}},
  \bibinfo {author} {\bibfnamefont {F.}~\bibnamefont {Lemmel}}, \bibinfo
  {author} {\bibfnamefont {P.-E.}\ \bibnamefont {Crouzet}}, \bibinfo {author}
  {\bibfnamefont {P.}~\bibnamefont {Verhoeve}},\ and\ \bibinfo {author}
  {\bibfnamefont {B.}~\bibnamefont {Shortt}},\ }\bibfield  {title} {\bibinfo
  {title} {Pyxel: a novel and multi-purpose python-based framework for imaging
  detector simulation},\ }in\ \href@noop {} {\emph {\bibinfo {booktitle} {High
  Energy, Optical, and Infrared Detectors for Astronomy VIII}}},\ Vol.\
  \bibinfo {volume} {10709}\ (\bibinfo {organization} {International Society
  for Optics and Photonics},\ \bibinfo {year} {2018})\ p.\ \bibinfo {pages}
  {107091A}\BibitemShut {NoStop}%
\bibitem [{\citenamefont {Donnelly}\ \emph {et~al.}(2020)\citenamefont
  {Donnelly}, \citenamefont {Finizio}, \citenamefont {Gliga}, \citenamefont
  {Holler}, \citenamefont {Hrabec}, \citenamefont {Odstr{\v{c}}il},
  \citenamefont {Mayr}, \citenamefont {Scagnoli}, \citenamefont {Heyderman},
  \citenamefont {Guizar-Sicairos} \emph {et~al.}}]{donnelly2020time}%
  \BibitemOpen
  \bibfield  {author} {\bibinfo {author} {\bibfnamefont {C.}~\bibnamefont
  {Donnelly}}, \bibinfo {author} {\bibfnamefont {S.}~\bibnamefont {Finizio}},
  \bibinfo {author} {\bibfnamefont {S.}~\bibnamefont {Gliga}}, \bibinfo
  {author} {\bibfnamefont {M.}~\bibnamefont {Holler}}, \bibinfo {author}
  {\bibfnamefont {A.}~\bibnamefont {Hrabec}}, \bibinfo {author} {\bibfnamefont
  {M.}~\bibnamefont {Odstr{\v{c}}il}}, \bibinfo {author} {\bibfnamefont
  {S.}~\bibnamefont {Mayr}}, \bibinfo {author} {\bibfnamefont {V.}~\bibnamefont
  {Scagnoli}}, \bibinfo {author} {\bibfnamefont {L.~J.}\ \bibnamefont
  {Heyderman}}, \bibinfo {author} {\bibfnamefont {M.}~\bibnamefont
  {Guizar-Sicairos}}, \emph {et~al.},\ }\bibfield  {title} {\bibinfo {title}
  {Time-resolved imaging of three-dimensional nanoscale magnetization
  dynamics},\ }\href@noop {} {\bibfield  {journal} {\bibinfo  {journal} {Nature
  Nanotechnology}\ }\textbf {\bibinfo {volume} {15}},\ \bibinfo {pages} {356}
  (\bibinfo {year} {2020})}\BibitemShut {NoStop}%
\bibitem [{\citenamefont {Donnelly}\ \emph {et~al.}(2018)\citenamefont
  {Donnelly}, \citenamefont {Gliga}, \citenamefont {Scagnoli}, \citenamefont
  {Holler}, \citenamefont {Raabe}, \citenamefont {Heyderman},\ and\
  \citenamefont {Guizar-Sicairos}}]{donnelly2018tomographic}%
  \BibitemOpen
  \bibfield  {author} {\bibinfo {author} {\bibfnamefont {C.}~\bibnamefont
  {Donnelly}}, \bibinfo {author} {\bibfnamefont {S.}~\bibnamefont {Gliga}},
  \bibinfo {author} {\bibfnamefont {V.}~\bibnamefont {Scagnoli}}, \bibinfo
  {author} {\bibfnamefont {M.}~\bibnamefont {Holler}}, \bibinfo {author}
  {\bibfnamefont {J.}~\bibnamefont {Raabe}}, \bibinfo {author} {\bibfnamefont
  {L.~J.}\ \bibnamefont {Heyderman}},\ and\ \bibinfo {author} {\bibfnamefont
  {M.}~\bibnamefont {Guizar-Sicairos}},\ }\bibfield  {title} {\bibinfo {title}
  {Tomographic reconstruction of a three-dimensional magnetization vector
  field},\ }\href@noop {} {\bibfield  {journal} {\bibinfo  {journal} {New
  Journal of Physics}\ }\textbf {\bibinfo {volume} {20}},\ \bibinfo {pages}
  {083009} (\bibinfo {year} {2018})}\BibitemShut {NoStop}%
\bibitem [{\citenamefont {Hierro-Rodriguez}\ \emph {et~al.}(2018)\citenamefont
  {Hierro-Rodriguez}, \citenamefont {G{\"u}rsoy}, \citenamefont {Phatak},
  \citenamefont {Quir{\'o}s}, \citenamefont {Sorrentino}, \citenamefont
  {{\'A}lvarez-Prado}, \citenamefont {V{\'e}lez}, \citenamefont {Mart{\'\i}n},
  \citenamefont {Alameda}, \citenamefont {Pereiro} \emph
  {et~al.}}]{hierro20183d}%
  \BibitemOpen
  \bibfield  {author} {\bibinfo {author} {\bibfnamefont {A.}~\bibnamefont
  {Hierro-Rodriguez}}, \bibinfo {author} {\bibfnamefont {D.}~\bibnamefont
  {G{\"u}rsoy}}, \bibinfo {author} {\bibfnamefont {C.}~\bibnamefont {Phatak}},
  \bibinfo {author} {\bibfnamefont {C.}~\bibnamefont {Quir{\'o}s}}, \bibinfo
  {author} {\bibfnamefont {A.}~\bibnamefont {Sorrentino}}, \bibinfo {author}
  {\bibfnamefont {L.~M.}\ \bibnamefont {{\'A}lvarez-Prado}}, \bibinfo {author}
  {\bibfnamefont {M.}~\bibnamefont {V{\'e}lez}}, \bibinfo {author}
  {\bibfnamefont {J.~I.}\ \bibnamefont {Mart{\'\i}n}}, \bibinfo {author}
  {\bibfnamefont {J.~M.}\ \bibnamefont {Alameda}}, \bibinfo {author}
  {\bibfnamefont {E.}~\bibnamefont {Pereiro}}, \emph {et~al.},\ }\bibfield
  {title} {\bibinfo {title} {3d reconstruction of magnetization from dichroic
  soft x-ray transmission tomography},\ }\href@noop {} {\bibfield  {journal}
  {\bibinfo  {journal} {Journal of synchrotron radiation}\ }\textbf {\bibinfo
  {volume} {25}},\ \bibinfo {pages} {1144} (\bibinfo {year}
  {2018})}\BibitemShut {NoStop}%
\bibitem [{\citenamefont {Grollier}\ \emph {et~al.}(2020)\citenamefont
  {Grollier}, \citenamefont {Querlioz}, \citenamefont {Camsari}, \citenamefont
  {Everschor-Sitte}, \citenamefont {Fukami},\ and\ \citenamefont
  {Stiles}}]{grollier2020neuromorphic}%
  \BibitemOpen
  \bibfield  {author} {\bibinfo {author} {\bibfnamefont {J.}~\bibnamefont
  {Grollier}}, \bibinfo {author} {\bibfnamefont {D.}~\bibnamefont {Querlioz}},
  \bibinfo {author} {\bibfnamefont {K.}~\bibnamefont {Camsari}}, \bibinfo
  {author} {\bibfnamefont {K.}~\bibnamefont {Everschor-Sitte}}, \bibinfo
  {author} {\bibfnamefont {S.}~\bibnamefont {Fukami}},\ and\ \bibinfo {author}
  {\bibfnamefont {M.~D.}\ \bibnamefont {Stiles}},\ }\bibfield  {title}
  {\bibinfo {title} {Neuromorphic spintronics},\ }\href@noop {} {\bibfield
  {journal} {\bibinfo  {journal} {Nature electronics}\ }\textbf {\bibinfo
  {volume} {3}},\ \bibinfo {pages} {360} (\bibinfo {year} {2020})}\BibitemShut
  {NoStop}%
\bibitem [{\citenamefont {Hayward}(2015)}]{hayward2015intrinsic}%
  \BibitemOpen
  \bibfield  {author} {\bibinfo {author} {\bibfnamefont {T.}~\bibnamefont
  {Hayward}},\ }\bibfield  {title} {\bibinfo {title} {Intrinsic nature of
  stochastic domain wall pinning phenomena in magnetic nanowire devices},\
  }\href@noop {} {\bibfield  {journal} {\bibinfo  {journal} {Scientific
  reports}\ }\textbf {\bibinfo {volume} {5}},\ \bibinfo {pages} {1} (\bibinfo
  {year} {2015})}\BibitemShut {NoStop}%
\bibitem [{\citenamefont {Kimel}\ \emph {et~al.}(2022)\citenamefont {Kimel},
  \citenamefont {Zvezdin}, \citenamefont {Sharma}, \citenamefont {Shallcross},
  \citenamefont {De~Sousa}, \citenamefont {Garcia-Martin}, \citenamefont
  {Salvan}, \citenamefont {Hamrle}, \citenamefont {Stejskal}, \citenamefont
  {Mccord} \emph {et~al.}}]{kimel20222022}%
  \BibitemOpen
  \bibfield  {author} {\bibinfo {author} {\bibfnamefont {A.}~\bibnamefont
  {Kimel}}, \bibinfo {author} {\bibfnamefont {A.}~\bibnamefont {Zvezdin}},
  \bibinfo {author} {\bibfnamefont {S.}~\bibnamefont {Sharma}}, \bibinfo
  {author} {\bibfnamefont {S.}~\bibnamefont {Shallcross}}, \bibinfo {author}
  {\bibfnamefont {N.}~\bibnamefont {De~Sousa}}, \bibinfo {author}
  {\bibfnamefont {A.}~\bibnamefont {Garcia-Martin}}, \bibinfo {author}
  {\bibfnamefont {G.}~\bibnamefont {Salvan}}, \bibinfo {author} {\bibfnamefont
  {J.}~\bibnamefont {Hamrle}}, \bibinfo {author} {\bibfnamefont
  {O.}~\bibnamefont {Stejskal}}, \bibinfo {author} {\bibfnamefont
  {J.}~\bibnamefont {Mccord}}, \emph {et~al.},\ }\bibfield  {title} {\bibinfo
  {title} {The 2022 magneto-optics roadmap},\ }\href@noop {} {\bibfield
  {journal} {\bibinfo  {journal} {Journal of Physics D: Applied Physics}\
  }\textbf {\bibinfo {volume} {55}},\ \bibinfo {pages} {463003} (\bibinfo
  {year} {2022})}\BibitemShut {NoStop}%
\end{thebibliography}%

\end{document}